\documentclass[preprint,12pt]{elsarticle}

\usepackage{amssymb}
\usepackage{float}
\usepackage{graphicx}
\usepackage{subfigure}
\usepackage{leftidx}
\biboptions{sort&compress}

\journal{Nuclear Instruments and Methods in Physics Research A}

\begin{document}

\begin{frontmatter}



\title{Neutron Beam Tests of $CsI(Na)$ and $CaF_{2}(Eu)$ Crystals for Dark Matter Direct Search}


\author{$C.~Guo^{a,b}$, $X.H.~Ma^{a}$, $Z.M.~Wang^{a,}$\footnote{Corresponding author at: Key Laboratory of Particle Astrophysics, Institute of High Energy Physics, Chinese Academy of Sciences, 19B Yuquan Road, Beijing 100049, China.
Tel:~+86-1088235437. E-mail address: wangzhm@ihep.ac.cn (Z.M.~Wang).}, $J.~Bao^{c}$, $C.J.~Dai^{a}$, $M.Y.~Guan^{a}$, $J.C.~Liu^{a}$, $Z.H.~Li^{a}$, $J.~Ren^{c}$, $X.C.~Ruan^{c}$, $C.G.~Yang^{a}$, $Z.Y.~Yu^{a}$, $W.L.~Zhong^{a}$, $C.~Huerta^{d}$}

\address{${^a}$Key Laboratory of Particle Astrophysics, Institute of High Energy Physics, Chinese Academy of Science,Beijing, China\\
${^b}$ School of Physics, University of Chinese Academy of Science, Beijing, China \\
${^c}$ Science and Technology on Nuclear Data Laboratory, China Institute of Atomic Energy, Beijing, China\\
${^d}$ Department of Physics, Northern Illinois University, DeKalb, IL, USA}

\begin{abstract}
In recent decades, inorganic crystals have been widely used in dark matter direct search experiments. To contribute to the understanding of the capabilities of $CsI(Na)$ and $CaF_{2}(Eu)$ crystals, a mono-energetic neutron beam is utilized to study the properties of nuclear recoils, which are expected to be similar to signals of dark matter direct detection. The quenching factor of nuclear recoils in $CsI(Na)$ and $CaF_{2}(Eu)$, as well as an improved discrimination factor between nuclear recoils and ${\gamma}$ backgrounds in $CsI(Na)$, are reported.

\end{abstract}

\begin{keyword}
Dark matter direct search\sep Neutron beam test\sep $CaF_{2}(Eu)$ \sep $CsI(Na)$ \sep Elastic scattering



\end{keyword}

\end{frontmatter}


\section{Introduction}
\par
According to the recent results of the Planck~\cite{Planck}, it is known that the normal matter constitutes only 4.9\% of the universe's mass/energy inventory. Dark matter, which is observed indirectly by its gravitational influence on nearby matter, occupies 26.8\%, while the dark energy, thought to be responsible for accelerating the expansion of the universe, accounts for 68.3\%. The Weakly Interacting Massive Particle (WIMP) is a popular dark matter candidate. To directly observe it, the most promising method is expected to be the detection of the nuclear recoil signals due to the elastic scattering between WIMPs and a target nuclei, for example in the inorganic crystals.
\par
In recent decades, neutron beam tests have been performed for various crystals, including $NaI(Tl)$~\cite{Spooner,Gerbier,Chagani,Collar,Tovey,Simon,Jagemann}, $CsI(Tl)$~\cite{Pecourt,Wang,KIMS}, $CsI(Na)$~\cite{KIMS,Xilei} and  $CaF_{2}$~\cite{Tovey,Hazama,Bacci}. According to Ref.\cite{Xilei}, $CsI(Na)$ may be a good candidate for dark matter direct detection because of its high neutron/${\gamma}$ discrimination ability, but not consistent with previous results~\cite{KIMS}; we report our results in this paper. On the other hand, $CaF_{2}(Eu)$ are sensitive to spin-dependent dark matter with the F content~\cite{Bacci,R.Bernabei,Ellis,Bernabei,Bottino}, but its neutron/${\gamma}$ discrimination ability is not well known which will be reported here.

\section{Experimental Setup}
\par
\begin{figure}[H]
\centering
\includegraphics[height=10cm,angle=270]{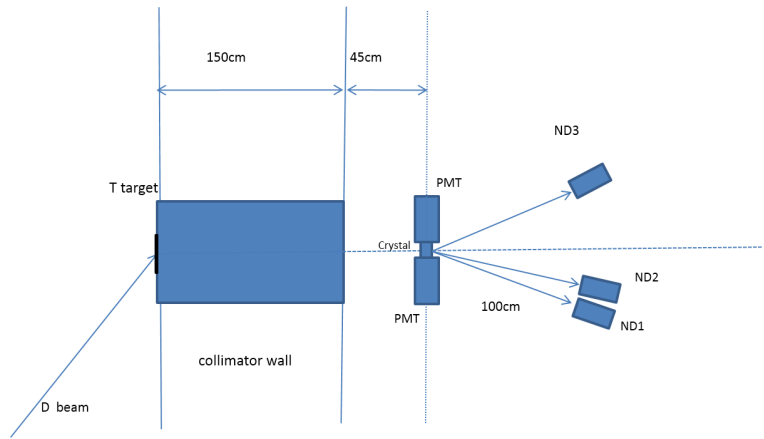}
\caption{The schematic diagram for the experiment}
\label{fig:schematic}
\end{figure}
The experiment is based on a neutron generator at the China Institute of Atomic Energy(CIAE) and the setup is shown in Fig.~\ref{fig:schematic}. The neutron beam is obtained from the T(D,n)${\alpha}$ reaction, which is induced by a 250~keV deuteron beam impinging on a T-Ti target with a frequency of 1.5~MHz and pulse width of 2~ns. Neutrons at an outgoing angle of 32.5~degrees are selected with a 1.5~m thick collimator wall, which is made of concrete, iron and lead, resulting a mean kinetic energy of 14.7~MeV with spread of 0.2~MeV (1${\sigma}$). The measured crystal is 0.45~m away from the wall and the neutron flux crossing the crystal is about 1${\times}10^{3}$/s/$cm^{2}$.
\begin{table}[H]
\begin{center}
\begin{tabular}[c]{ccccc} \hline
  & ND1 & ND2 & ND3\\\hline\hline
Degree-$CsI(Na)$ & $30^{\circ}{\pm}1^{\circ}$ & $50^{\circ}{\pm}1^{\circ}$ & $25^{\circ}{\pm}1^{\circ}$\\
$E_{recoil}$-Cs & $30.4^{+2.1}_{-1.9}$keV & $80.3^{+2.4}_{-2.0}$keV  & $21.2^{+1.7}_{-1.6}$keV\\
$E_{recoil}$-I  & $31.8^{+2.3}_{-2.0}$keV & $85.5^{+3.0}_{-3.2}$keV & $22.3^{+1.9}_{-1.8}$keV\\
Degree-$CaF_{2}(Eu)$ & $30^{\circ}{\pm}1^{\circ}$ &$15^{\circ}{\pm}1^{\circ}$ &$25^{\circ}{\pm}1^{\circ}$\\
$E_{recoil}$-Ca & $101.5^{+6.8}_{-7.0}$keV & $24.8^{+3.5}_{-3.7}$keV  & $70.4^{+6.3}_{-5.5}$keV\\
$E_{recoil}$-F & $212.1^{+13.5}_{-13.8}$keV & $53.4^{+8.0}_{-6.5}$keV & $148.0^{+13.3}_{-11.1}$keV\\\hline
\end{tabular}
\caption{Estimated recoiling energy of crystal samples}
\label{tab:crystal samples}
\end{center}
\end{table}
Two crystals, with all surfaces polished, are both 2.5${\times}$2.5${\times}$2.5$cm^{3}$ in cubic shape and produced by the Beijing Glass Research Institute. The doping concentrations of ${Na}$ in ${CsI(Na)}$ and ${Eu}$ in $CaF_{2}(Eu)$ are both 0.02\%. Two photomultipliers(PMTs) directly face the top and bottom surfaces of the crystal, while the other four surfaces are wrapped by a 65~${\mu}$m thick Enhanced Specular Reflector film. The PMTs, 9821QB from ET company, have a very low radioactivity background quartz window~\cite{ET}, and thus are particularly suitable for the future dark matter direct detection experiments.
\par
In order to select nuclear recoils with certain energies in the crystal, three neutron detectors(ND) are positioned at various angles and 1~m away from the crystal. The neutron detectors are made of liquid scintillator(BC501A), which has a good neutron/${\gamma}$ discrimination ability~\cite{Knoll}, contained in a cylindrical aluminum container of 5~cm diameter. Double checks with the laser alignment and the protractor ensure the setup height uncertainty to be less than 3~mm and positioning uncertainty less than 1 degree , then the uncertainties of angles are estimated as 1 degree. For each ND, a 2-inch XP2020 PMT is used for readout. The nuclear recoil energy($E_{recoil}$) can be calculated by the kinematical equation with the energy and scattering angle of the neutron:
\begin{eqnarray}
E_{recoil} = E_{beam} \{ 1-(\frac{m_{n}cos \theta -\sqrt{m^{2}_{N}+m^{2}_{n}sin^{2}\theta} }{m_{n}+m_{N}})^{2} \}
\label{Eq:recoiling energy}
\end{eqnarray}
where $E_{beam}$ is the neutron beam energy, $m_{n}$ and $m_{N}$ are the masses of a neutron and the recoiling nucleus(Cs, I, Ca, or F) respectively and ${\theta}$ is scattering angle. Table~\ref{tab:crystal samples} details the information of scattering angles and calculated recoil energies where the recoil energy uncertainties are propagated from the NDs' position uncertainties and neutron beam uncertainties.
\begin{figure}[H]
\centering
\includegraphics[height=10cm,angle=270]{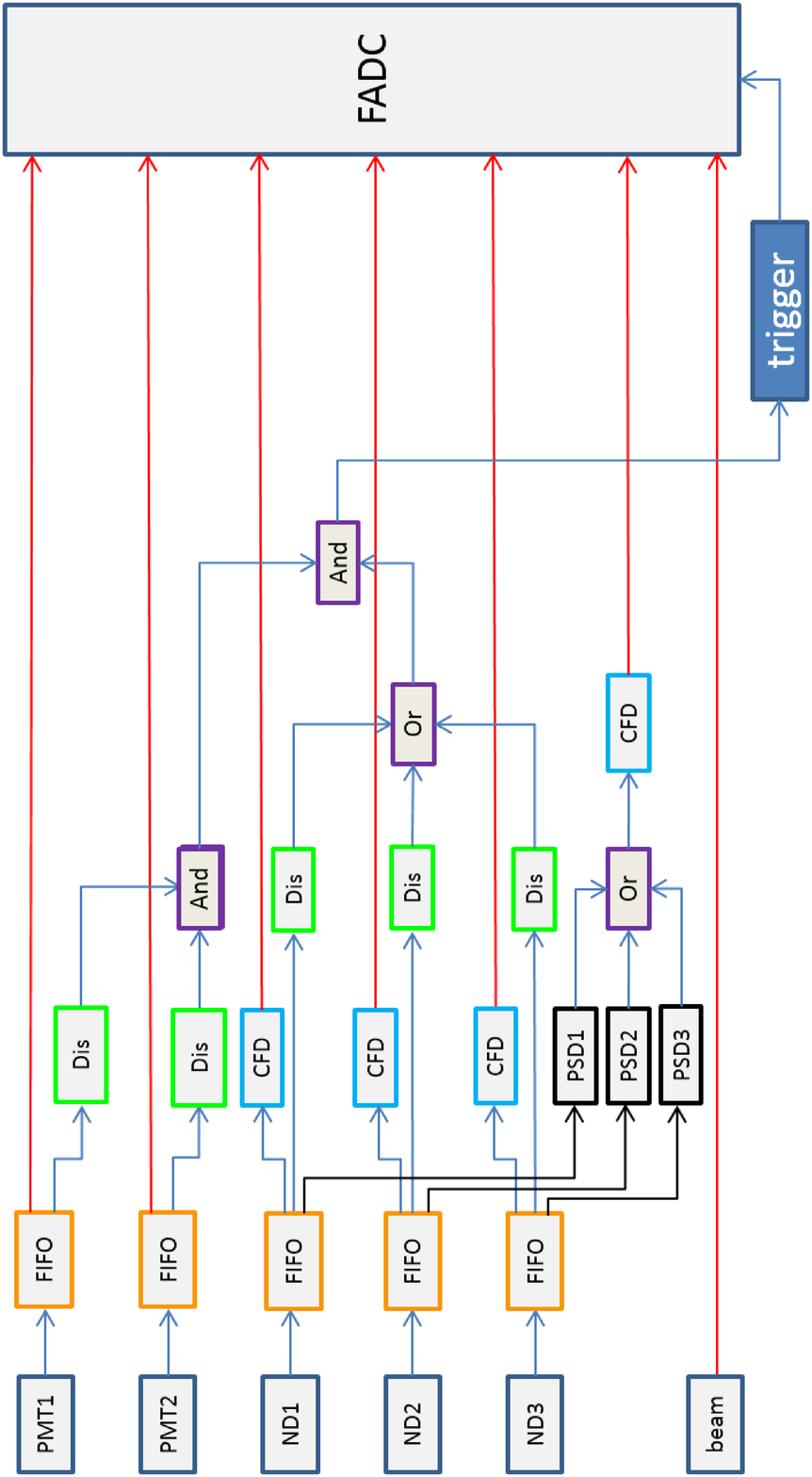}
\caption{Readout diagram of beam test}
\label{fig.readout}
\end{figure}
\par
Fig.~\ref{fig.readout} shows the electronics scheme of the experiment. In total there are 7 signal channels: 2 PMTs from the crystal detector named as PMT1 and PMT2; 3 PMTs from 3 NDs named as ND1, ND2 and ND3; a pulse shape discrimination (PSD) signal (generated by CANBERRA 2160A) from the 3 NDs; and a time stamp pulse from the neutron beam. The data is recorded by the Flash Analog to Digital Convertor (FADC CAEN V1729A, 2~GHz sampling frequency, 1.25~${\mu}$s readout window). The trigger, which is from the coincidence of the crystal detector and NDs for background suppressing, is about 1~Hz when the neutron beam is on. The sub-trigger of the crystal detector is generated by the coincidence of PMT1 and PMT2, where the single channel threshold is about 0.5~p.e..
\par
\begin{figure}[htb]
\centering
\subfigure{
\label{calibration result of CsI}
\includegraphics[width=4.3cm,height=6.5cm,angle=270]{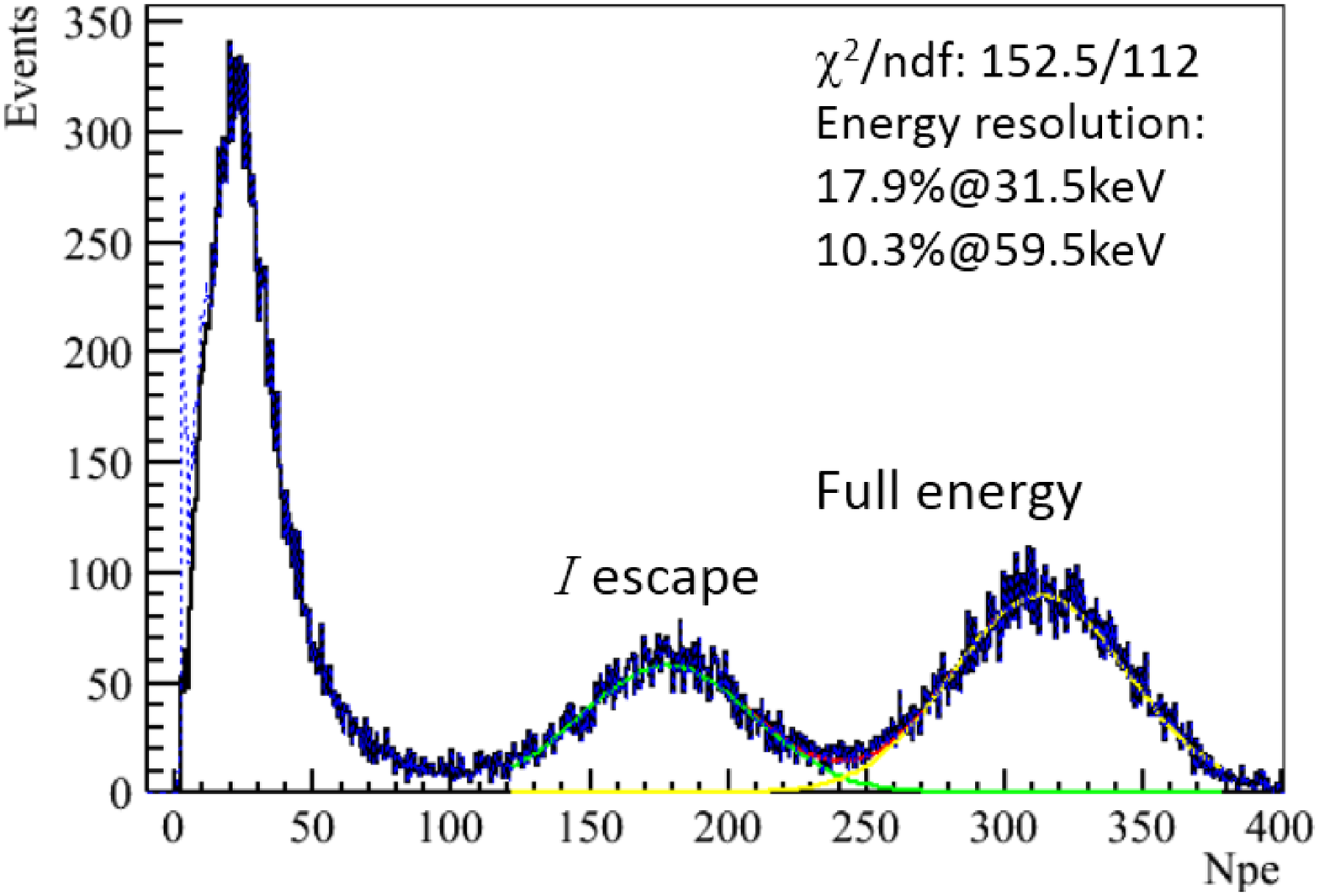}}
\subfigure{
\label{calibration result of CaF2}
\includegraphics[width=4.3cm,height=6.5cm,angle=270]{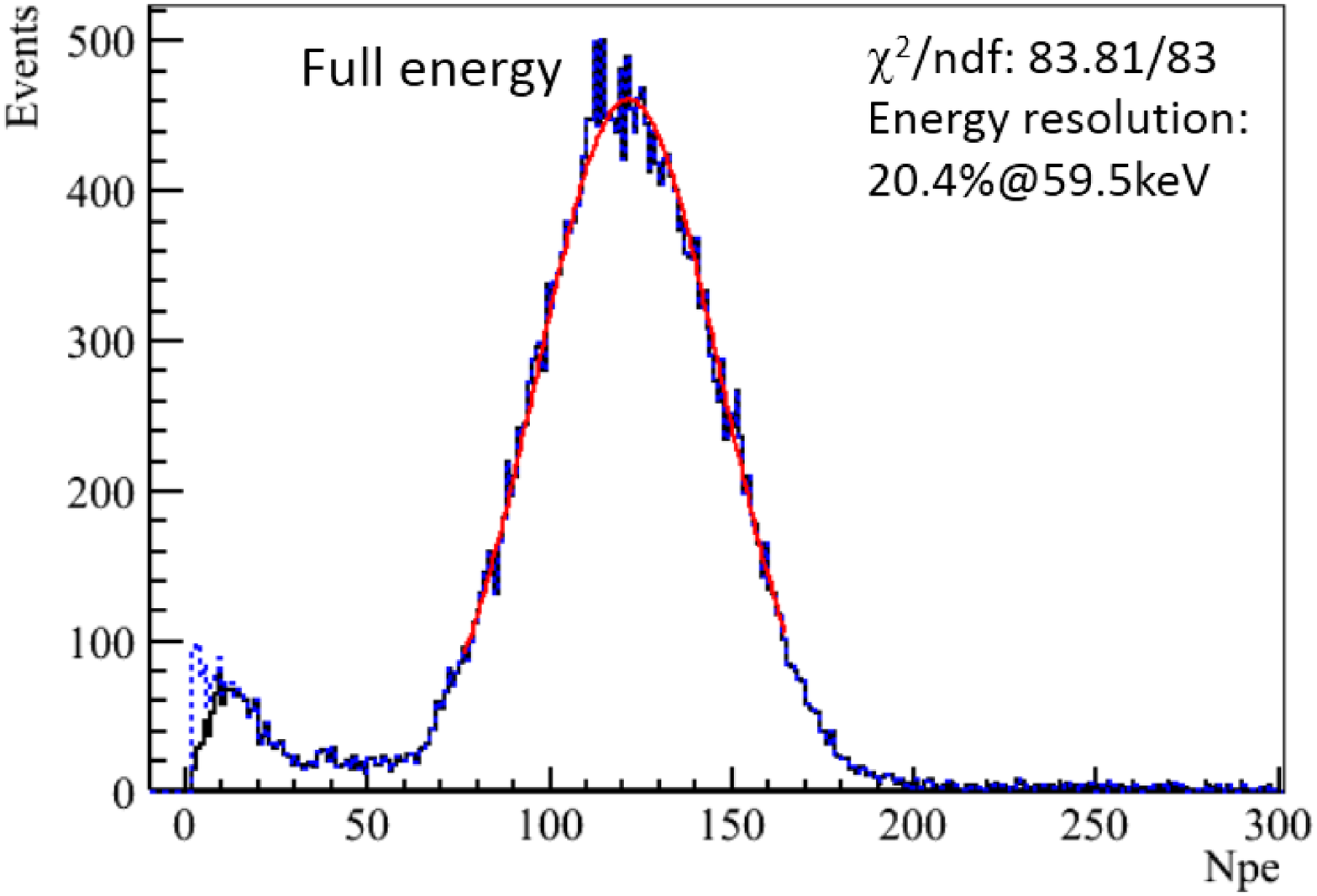}}
\caption{Calibration result with $^{241}Am$ ${\gamma}$ source. Left:$CsI(Na)$. Right:$CaF_{2}(Eu)$. The energy resolution is calculated with ${\sigma}/Mean$ of the fitting result. The blue dash lines are the spectra after trigger efficiency correction.}
\label{calibration result}
\end{figure}

The detectors are calibrated with ${\gamma}$ sources, $^{241}Am$ and ${^{137}Cs}$. The $^{241}Am$ ${\gamma}$ spectra of ${CsI(Na)}$ and ${CaF_{2}(Eu)}$ are shown in Fig.~\ref{calibration result}. The measured light yields of $CsI(Na)$ and $CaF_{2}(Eu)$ are 5.6 p.e/keV and 2.0 p.e./keV respectively. The effective trigger threshold of the crystal detectors is about 5~p.e.(50\% efficiency) which is mainly related to the crystal light emitting time constant and coincidence window, and its efficiency is checked by calibration sources, background and Toy Monte Carlo results, which induced uncertainty will be considered in the following analysis. While NDs are calibrated with $^{137}Cs$ and their threshold is 0.1~MeV.

\section{Data analysis}
\par
\begin{figure}[H]
\centering
\subfigure{
\label{CsI_tof_ND3}
\includegraphics[width=4.3cm,height=6.5cm,angle=270]{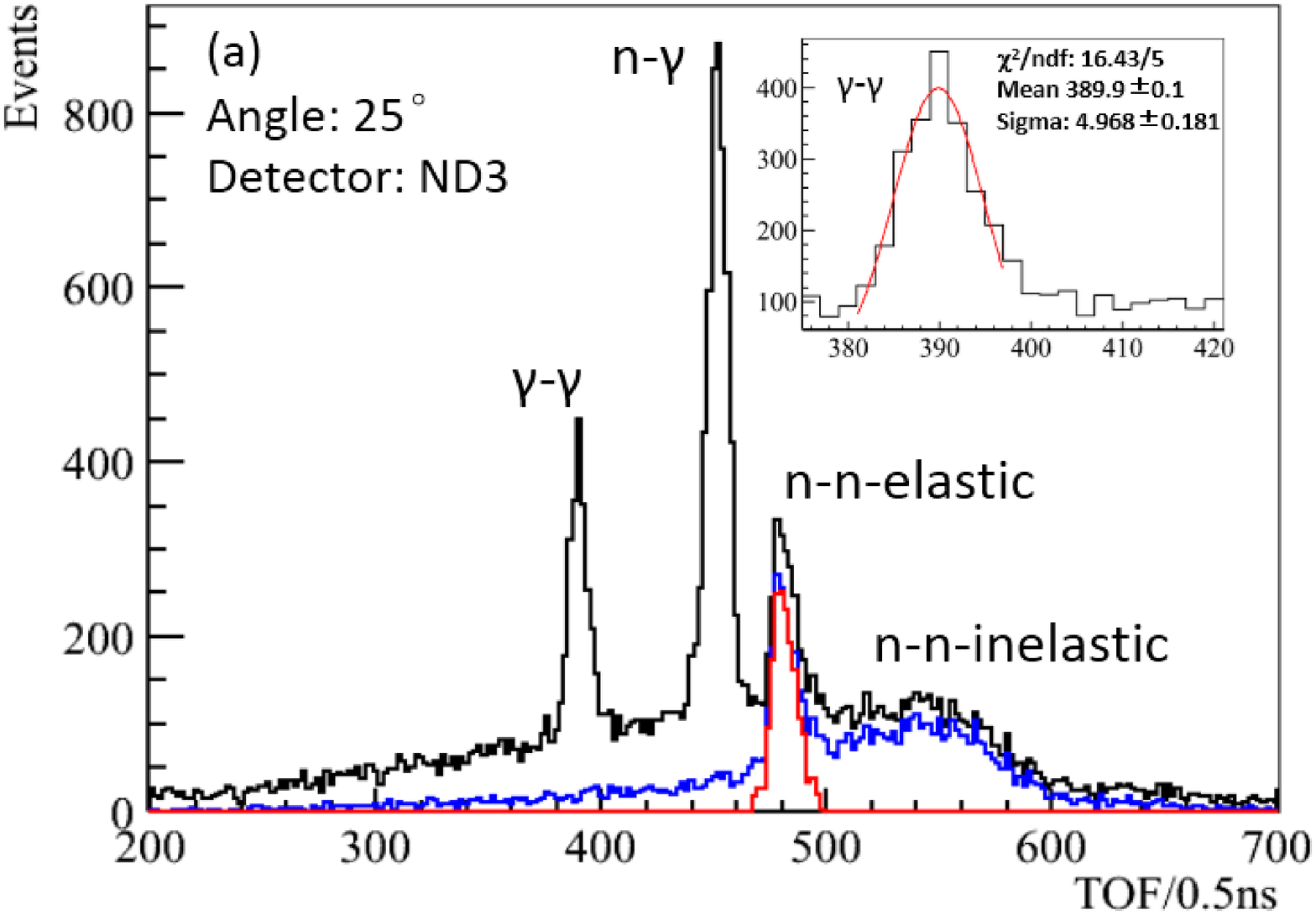}}
\subfigure{
\label{CsI_tof_ND1}
\includegraphics[width=4.3cm,height=6.5cm,angle=270]{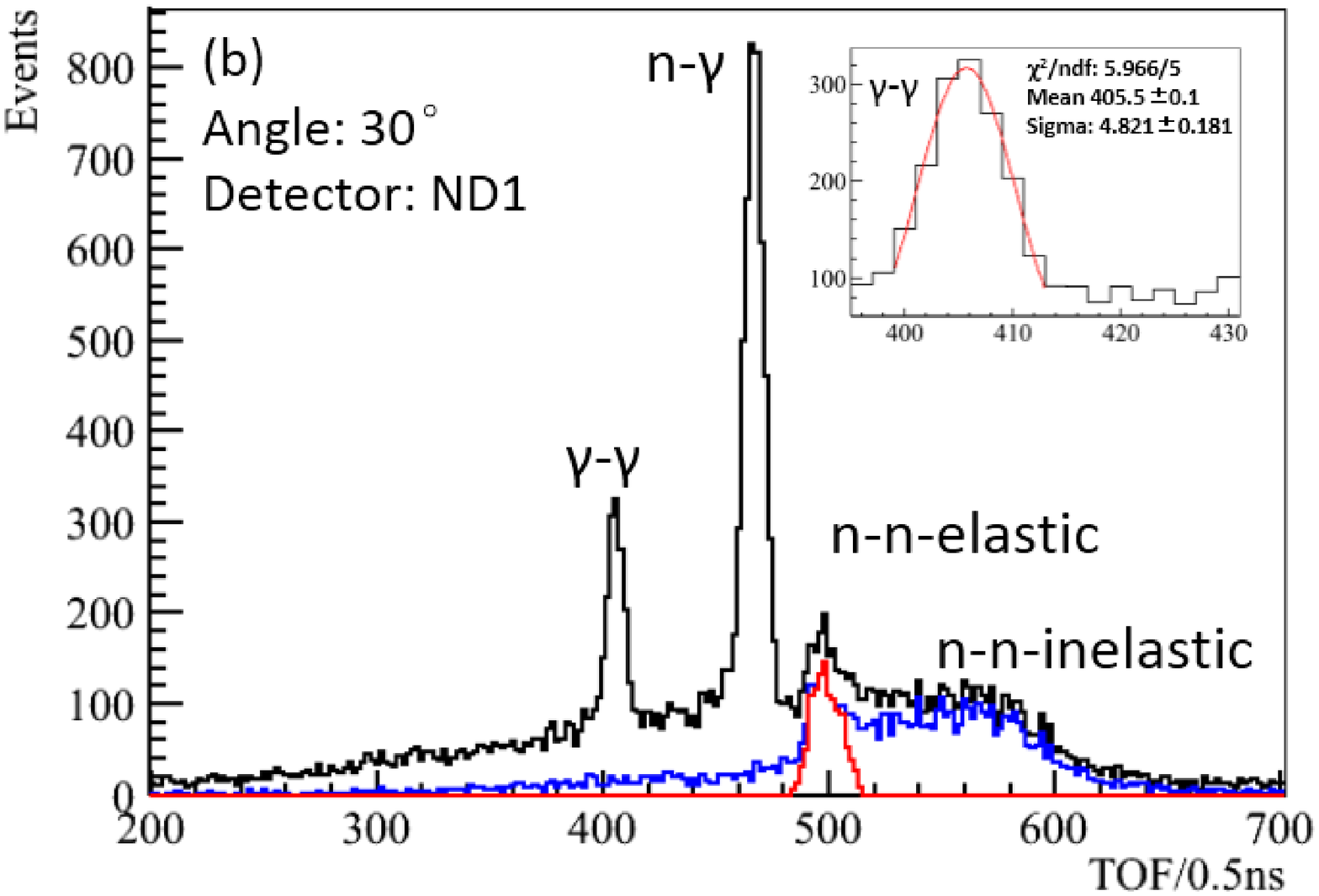}}
\subfigure{
\label{CsI_tof_ND2}
\includegraphics[width=4.3cm,height=6.5cm,angle=270]{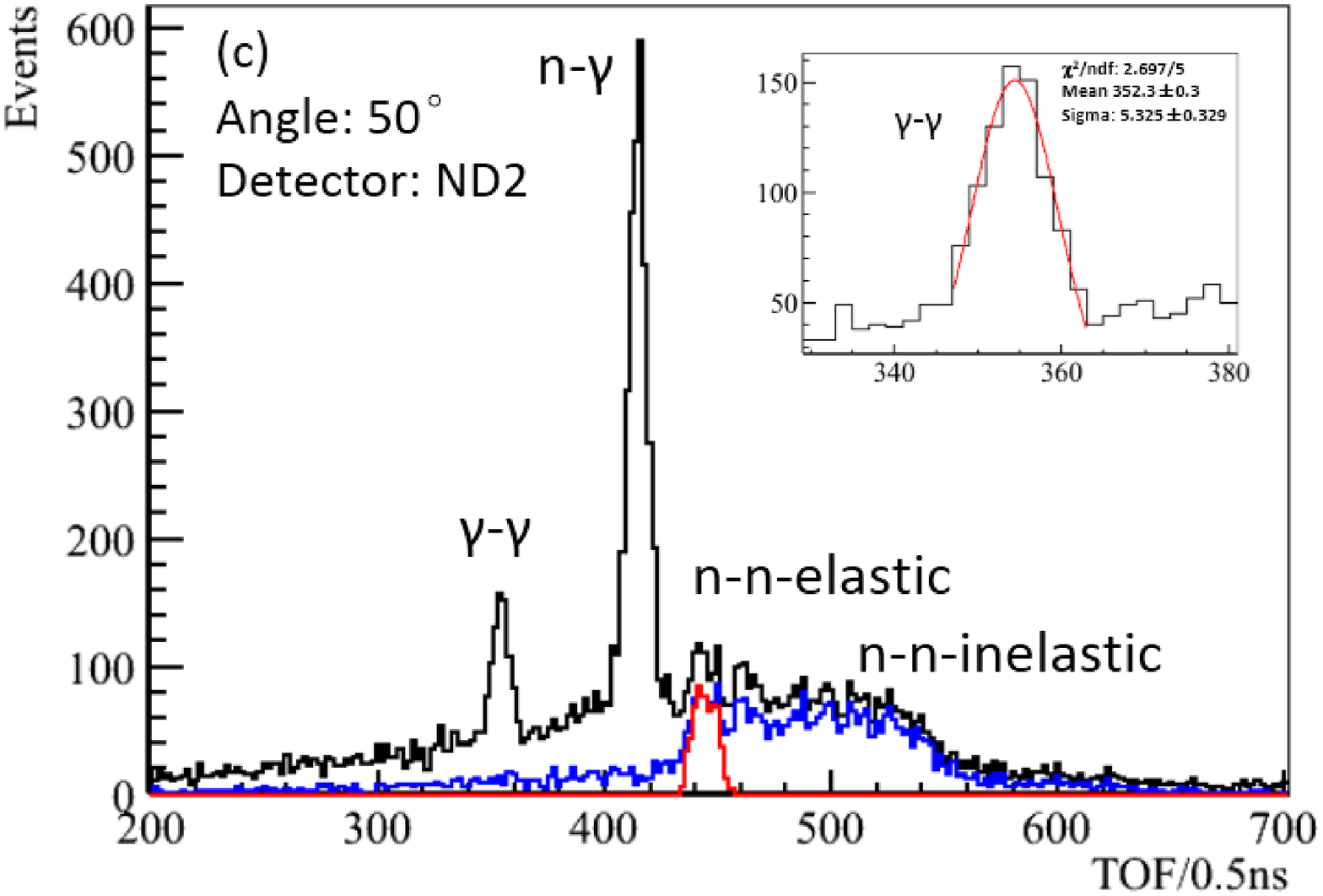}}
\caption{TOF distributions of $CsI(Na)$ from neutron source to NDs. Black line: Raw data. Blue line: After 2-D cut on ND energy and PSD (Fig.~\ref{Fig.PSD}). Red line: After TOF cut(Fig.~\ref{Fig.energy_tof.eps}).}
\label{Fig.tof_cut1}
\end{figure}

Neutron/${\gamma}$ discrimination is key to the analysis. In this section, the analysis procedures of $CsI(Na)$ are described in details as an example. The fluctuation of time of flight (TOF), one important input, is about 2.5~ns, which is calculated by fitting the ${\gamma}$-${\gamma}$ peak in Fig.~\ref{Fig.tof_cut1}.
\par
TOF distributions of $CsI(Na)$ (Fig.~\ref{Fig.tof_cut1}, black line) clearly have four peaks from left to right:\\\indent
1. The ${\gamma}$-${\gamma}$ peak: The peak is formed by the ${\gamma}$ generated along with the neutron beam, scattering from the crystal and triggering the NDs. Since ${\gamma}$ has the highest and fixed speed, this peak is on the far left and has the narrowest width.\\\indent
2. The n-${\gamma}$ peak: Neutrons react with the $CsI(Na)$ crystal via inelastic processes and the secondary ${\gamma}$ triggers NDs.\\\indent
3. The n-n-elastic peak: Neutrons elastically scatter with nucleus in the crystal then trigger NDs. Since elastic scattering is mono-energetic at the fixed scattering angle, this peak has a narrow width. \\\indent
4. The n-n-inelastic peak: Neutrons react with the crystal via inelastic processes, for example Cs(n,n${\gamma}$)Cs, and the neutrons trigger NDs. Because the inelastic scattered neutrons are not mono-energetic and the energy loss is higher than the elastic scattered ones, this peak is on the far right and has the widest distribution.\\\indent
Aside from the four peaks, the wide and nearly flat part in Fig.~\ref{Fig.tof_cut1} is due to the direct current component of the pulsed beam and the accidental coincidence of the neutrons from the room scattering.

\begin{figure}[H]
\centering
\includegraphics[width=4.3cm,height=6.5cm,angle=270]{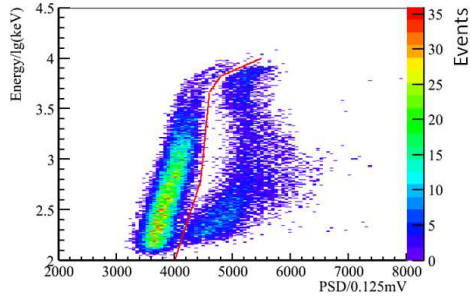}
\caption{Distribution of neutron energy deposited in ND3 vs PSD signal amplitude of $CsI(Na)$. The red line is the discrimination between neutron events on the right and ${\gamma}$ events on the left.}
\label{Fig.PSD}
\end{figure}
\par
To select clean scattering neutron samples, a 2-D cut on ND energy and PSD is applied(Fig.~\ref{Fig.PSD}). Then the neutron events, including elastic and inelastic neutrons, are clearly selected (Fig.~\ref{Fig.tof_cut1}, the blue line). The 2-D cut, compared with the traditional 1-D PSD cut, has higher background rejecting efficiency.

\par
\begin{figure}[htb]
\centering
\includegraphics[width=4.3cm,height=6.5cm,angle=270]{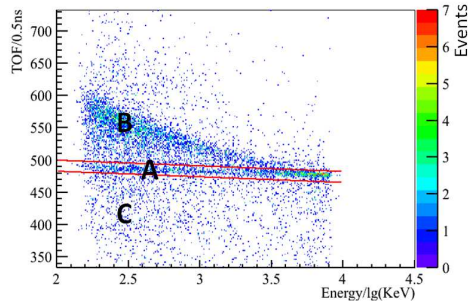}
\caption{Distribution of TOF vs neutron energy deposited in ND3. Part A: between the two red lines, elastic scattering events; Part B: inelastic scattering events; Part C: random coincident events.}
\label{Fig.energy_tof.eps}
\end{figure}

To further select the elastic scattering neutron events, a TOF cut is utilized as shown in Fig.~\ref{Fig.energy_tof.eps}, region A. Region B is the inelastic scattering events and region C is the accidental coincidence events. The cut is determined with a mean value predicted by the ${\gamma}$-${\gamma}$ peak plus elastic scattering neutron TOF and ${\pm}5$~ns (2 times of the fluctuation). The prediction is consistent with the data.

\begin{figure}[htb]
\centering
\subfigure{
\label{CsI_spectrum_ND3}
\includegraphics[width=4.3cm,height=6.5cm,angle=270]{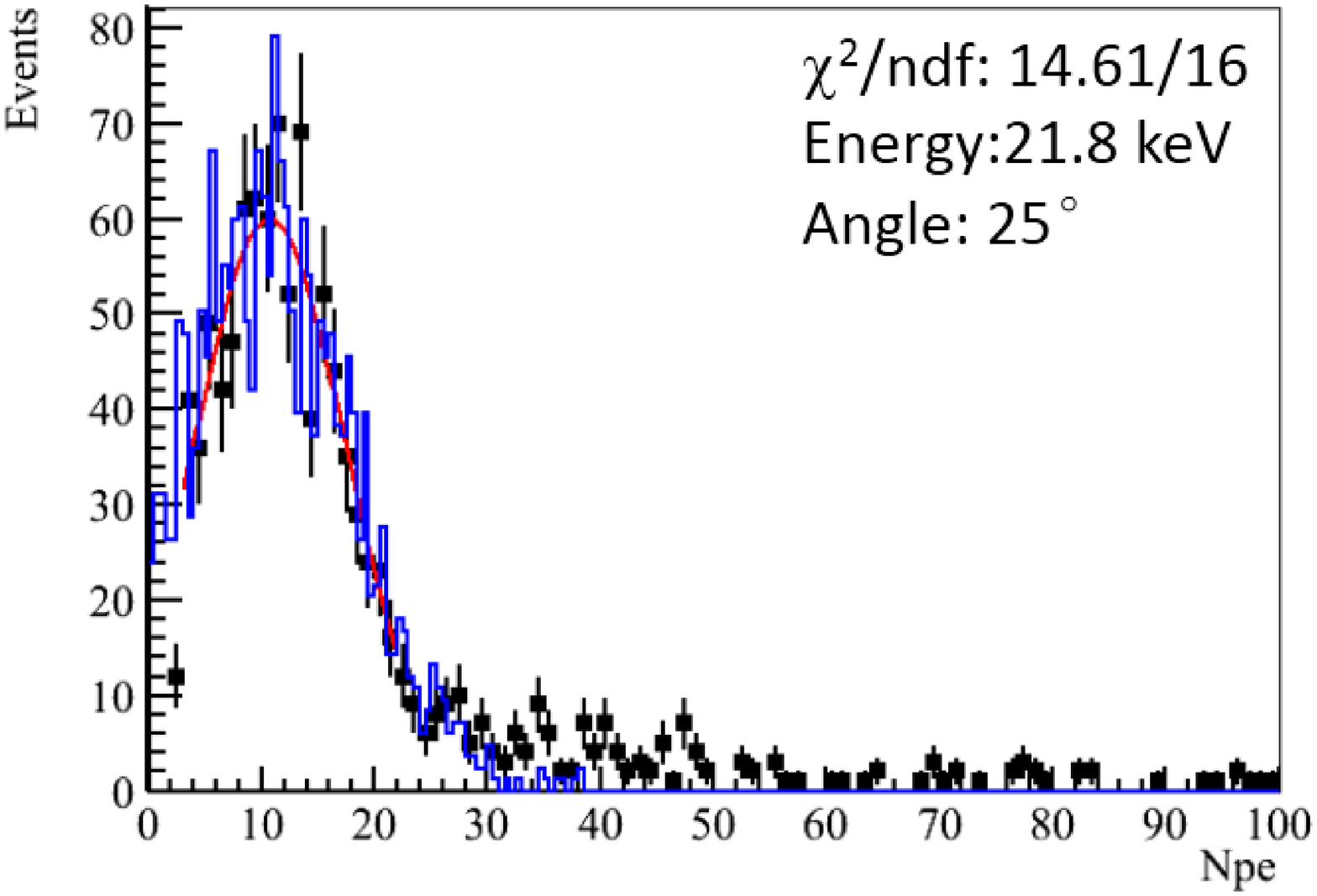}}
\subfigure{
\label{CsI_spectrum_ND1}
\includegraphics[width=4.3cm,height=6.5cm,angle=270]{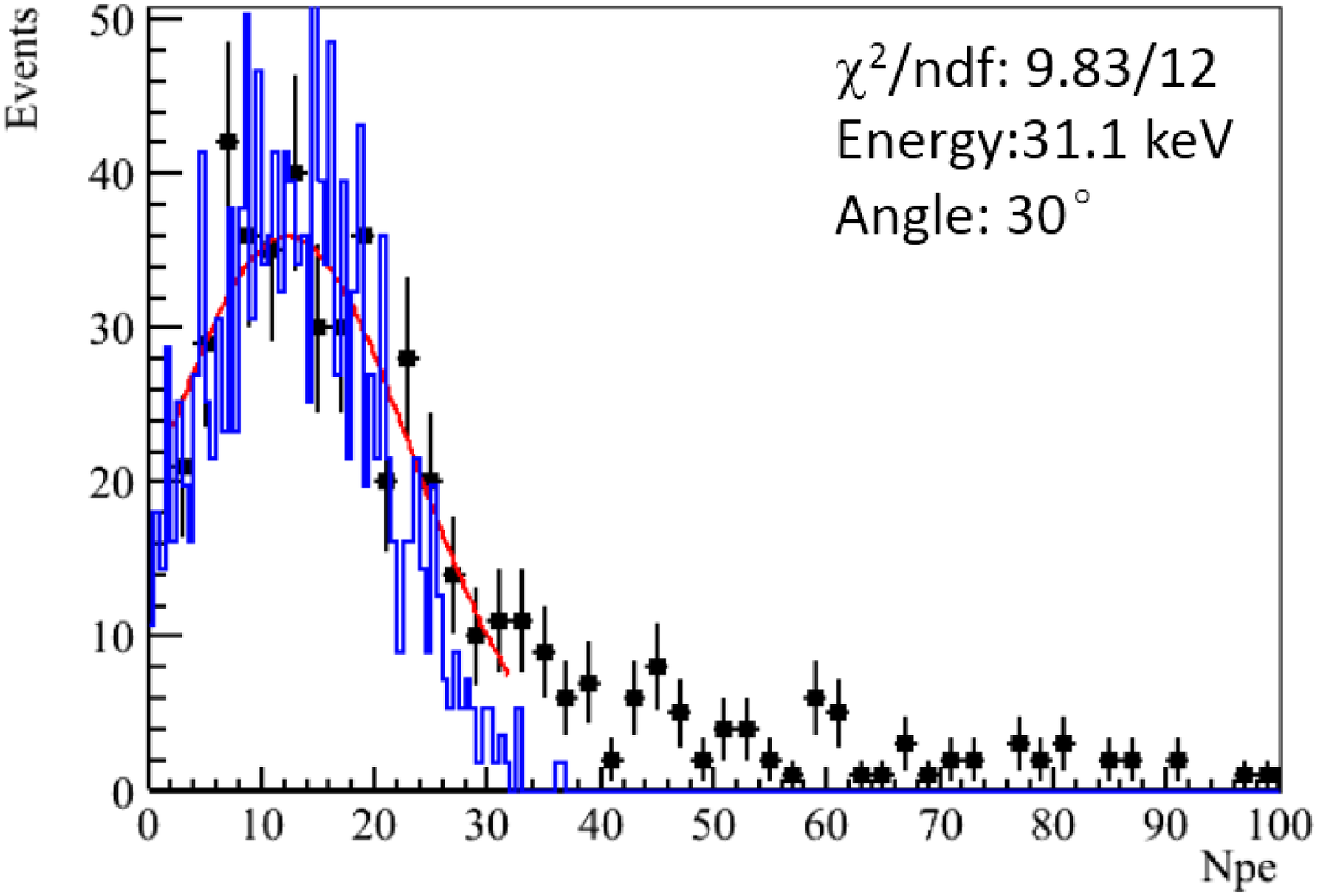}}
\subfigure{
\label{CsI_spectrum_ND2}
\includegraphics[width=4.3cm,height=6.5cm,angle=270]{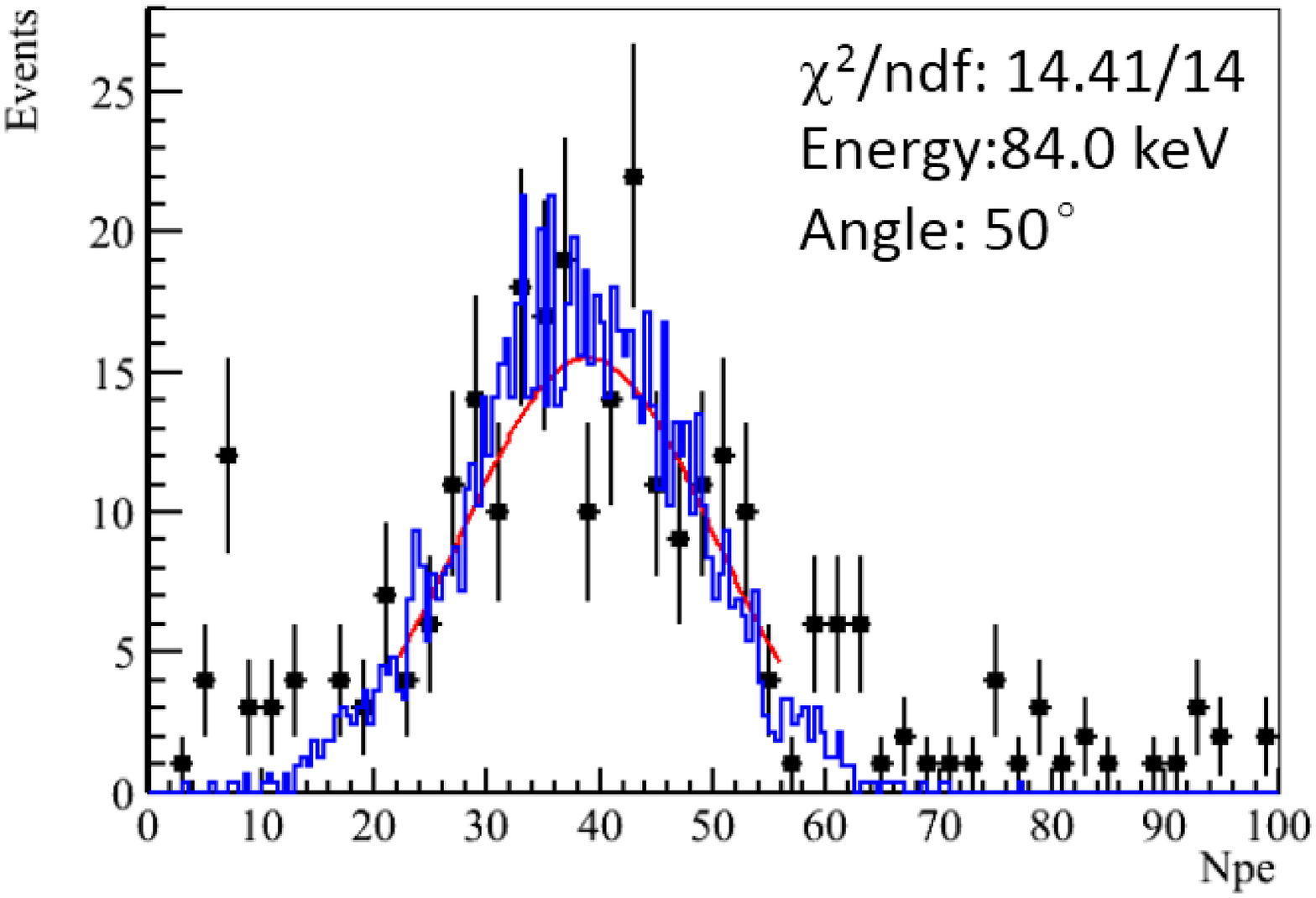}}
\caption{Recoil energy spectra of $CsI(Na)$ tagged by each ND and fitted with Gaussian function. The labeled energies are the average of Cs and I recoil energies. Black dots are experimental data and blue lines are Monte Carlo simulations.}
\label{Fig.spectrum of CsI(Na)}
\end{figure}

\par
\begin{figure}[H]
\centering
\subfigure{
\label{CaF2_spectrum_ND2}
\includegraphics[width=4.3cm,height=6.5cm,angle=270]{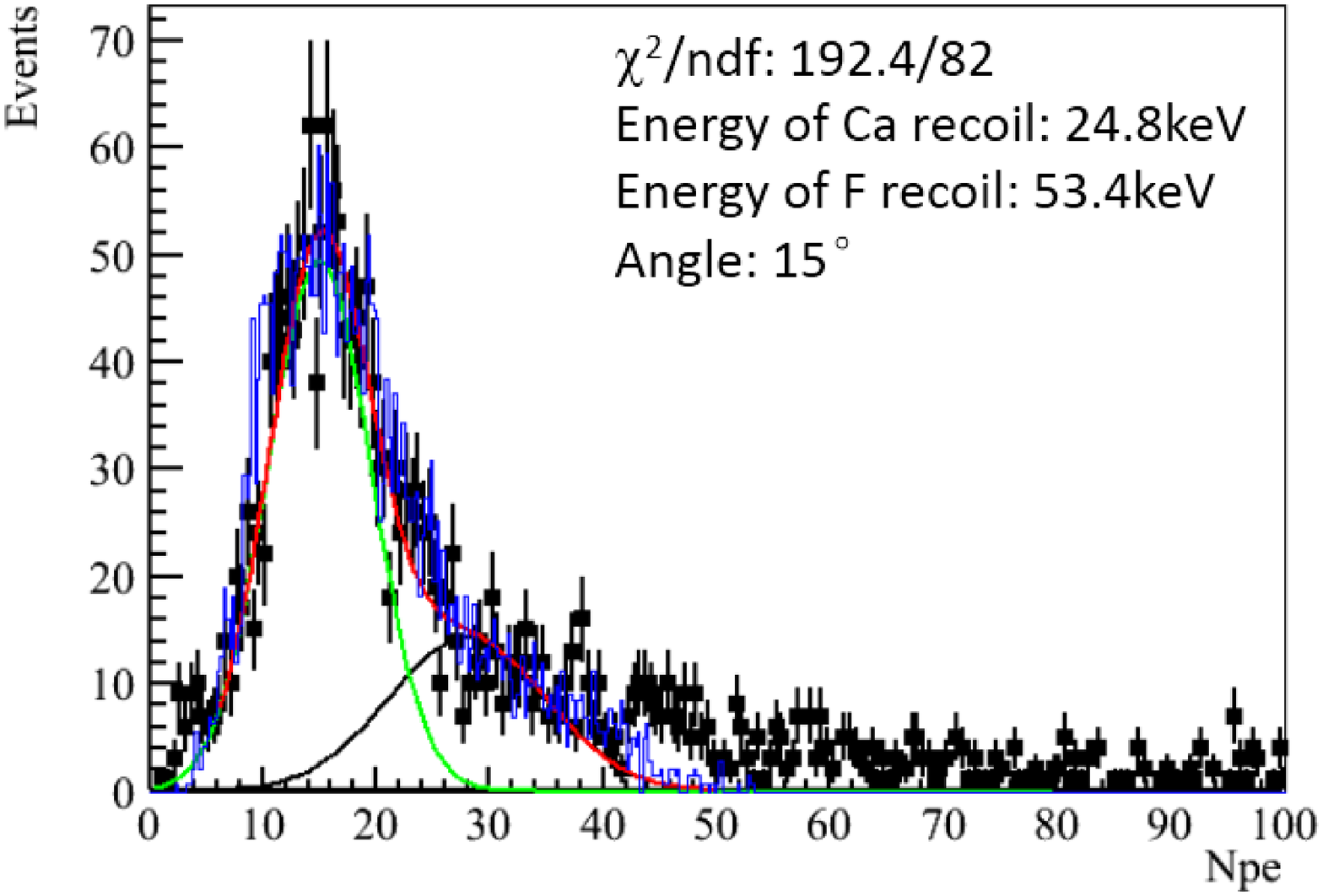}}
\subfigure{
\label{CaF2_spectrum_ND3}
\includegraphics[width=4.3cm,height=6.5cm,angle=270]{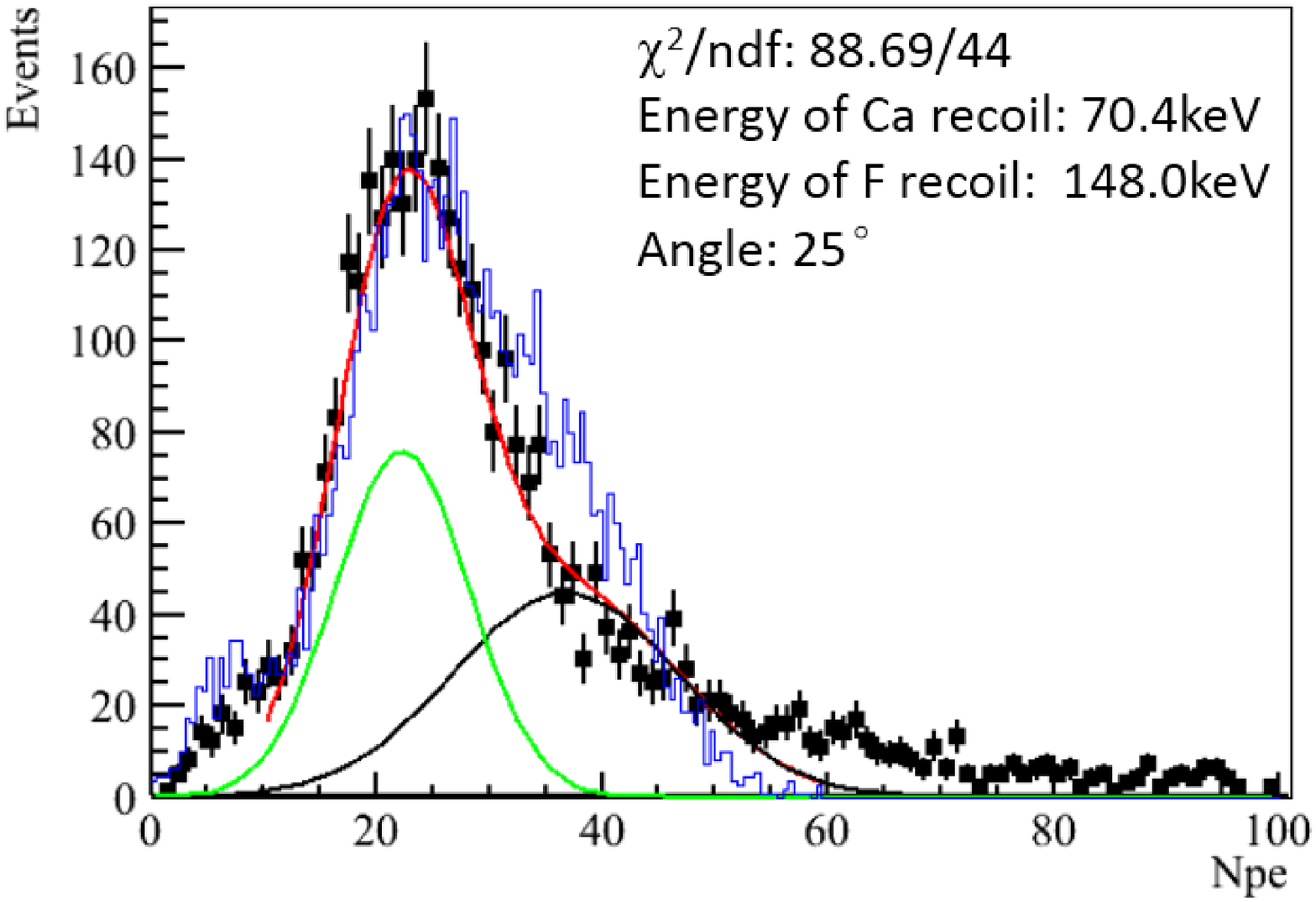}}
\subfigure{
\label{CaF2_spectrum_ND1}
\includegraphics[width=4.3cm,height=6.5cm,angle=270]{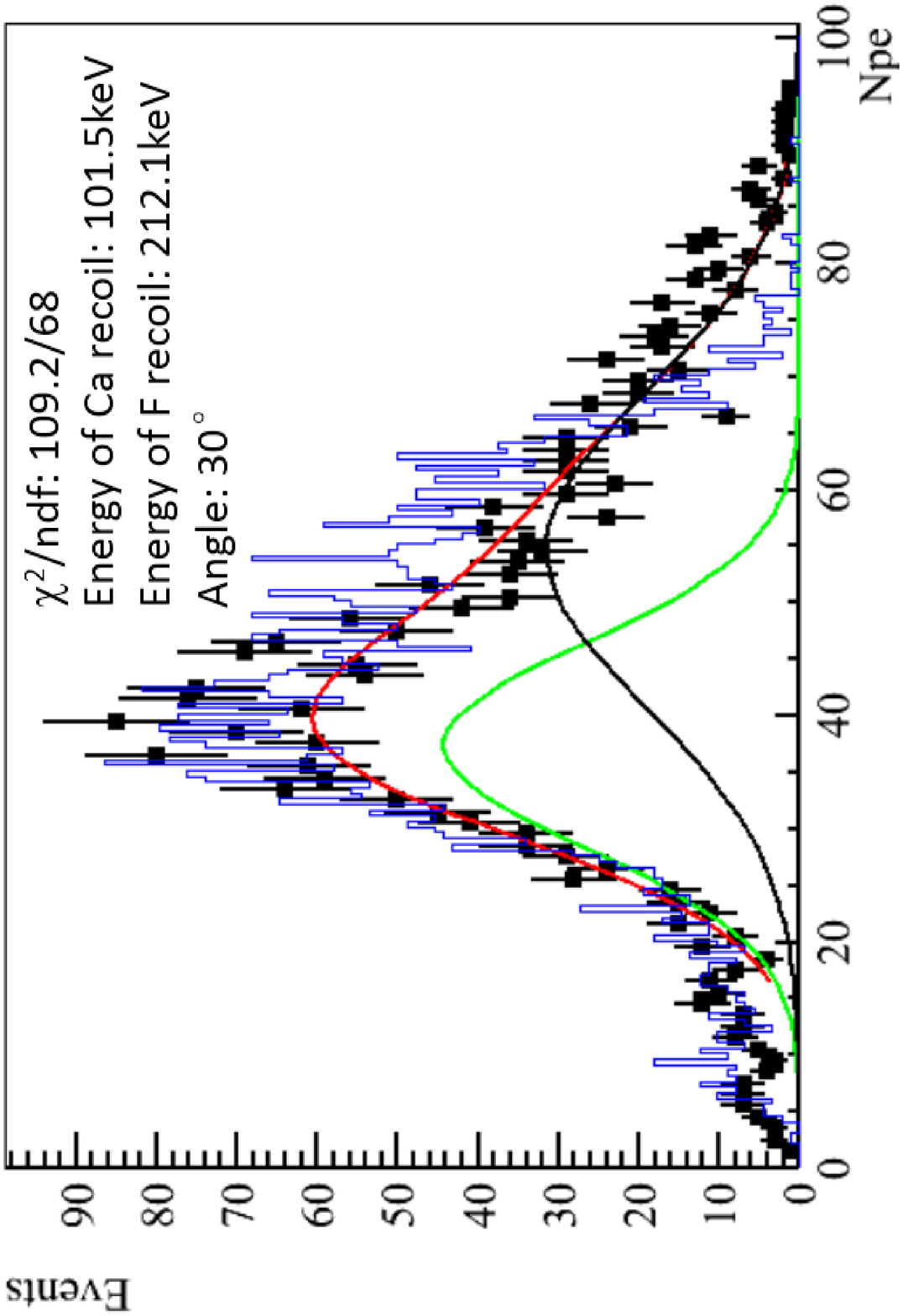}}
\caption{Energy spectra of $CaF_{2}(Eu)$ tagged by each ND and fitted with double Gaussian function. Black dots are experimental data and blue lines are Monte Carlo simulations.}
\label{Fig.spectrum of CaF2}
\end{figure}

The elastic scattering neutron events are clearly selected (Fig.~\ref{Fig.tof_cut1}, the red line), and number of photoelectron(Npe) distributions of $CsI(Na)$ are shown in Fig.~\ref{Fig.spectrum of CsI(Na)}. A Toy Monte Carlo is constructed to calculate the elastic scattering neutrons, including the effects of beam energy smear, detectors' geometry, NDs' efficiency and crystal response. The elastic scattering cross sections between nucleus and neutron are obtained from the National Nuclear Data Center database~\cite{NNDC}. The simulation results are also shown in Fig.~\ref{Fig.spectrum of CsI(Na)} and Fig.~\ref{Fig.spectrum of CaF2}, which are basically consistent with the data.  Because of the close mass of Cs and I, recoils from Cs or I could not be distinguished and the fitted results correspond to their averaged energy. For $CaF_{2}(Eu)$, the same analysis method is taken and the spectrum can be fitted with a double-Gaussian function for Ca and F recoiling(Fig.~\ref{Fig.spectrum of CaF2}).

\section{Results and discussion}
\subsection{Quenching factor}
\par
\begin{figure}[H]
\centering
\subfigure{
\label{fig:quenchingfactor for CsI}
\includegraphics[width=6.5cm,height=4.3cm]{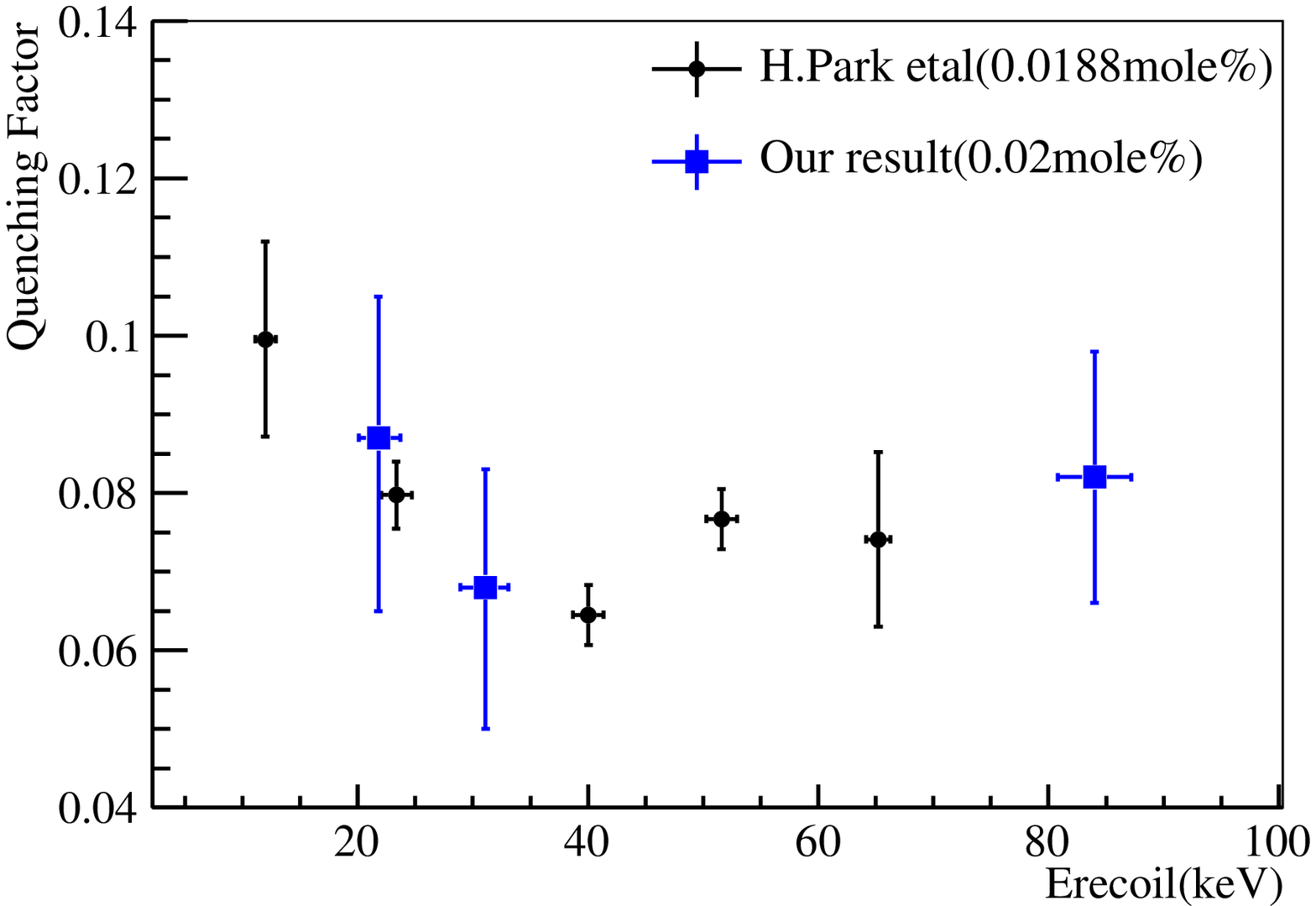}}
\subfigure{
\label{fig:quenching factor for CaF2}
\includegraphics[width=6.5cm,height=4.3cm]{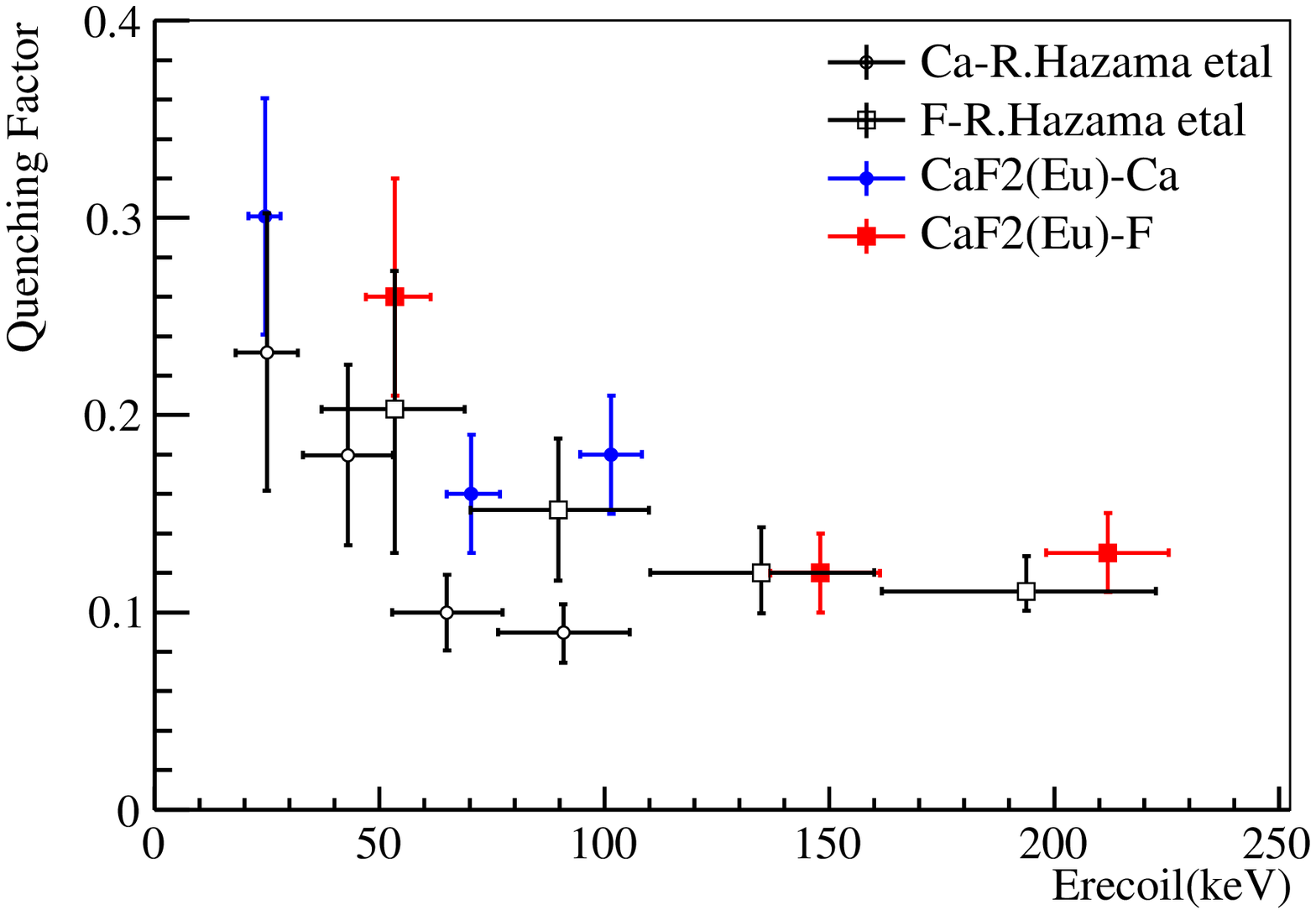}}
\caption{Quenching factors for crystal samples. Left:$CsI(Na)$. Right: $CaF_{2}(Eu)$. The present results are compared with H.Park etal~\cite{KIMS} and R.Hazama etal~\cite{Hazama}. The systematical and statistical errors are both included for R.Hazama and ours, while only statistical error in H.Park's results.}
\label{quenching factor for crystal samples}
\end{figure}

In a scintillation detector, organic and inorganic, energy of heavy ions always quenches and only part of its energy is released with scintillation photons. This fraction is called the quenching factor and it is an important property of the crystal. Generally speaking, larger value of quenching factor, i.e. higher light yield, results a better neutron/${\gamma}$ discrimination. The quenching factor is defined as
\begin{eqnarray}
Q=\frac{E_{meas}}{E_{recoil}}
\label{Eq:quenching}
\end{eqnarray}
where $E_{meas}$ is calculated with the measured p.e. normalized by the crystal light yield which is determined by ${^{241}Am}$ calibration data. $E_{recoil}$ is the recoil energy calculated with Eq.~\ref{Eq:recoiling energy}. For $CsI(Na)$, $E_{recoil}$ is the averaged recoil energy of Cs and I. The quenching factors of $CsI(Na)$ and $CaF_{2}(Eu)$(Fig.~\ref{quenching factor for crystal samples}) are consistent with the previous measurements~\cite{KIMS,Hazama} within uncertainties.

\par
In Fig.~\ref{quenching factor for crystal samples}, the horizontal uncertainties are dominated by the 1 degree scattering angle uncertainties and the vertical one includes the contribution of the statistics, trigger efficiency and the systematics, which is dominated by crystal response non-linearity to electrons. In Eq.~\ref{Eq:quenching}, it is assumed that the light yield is linear at different ${\gamma}$ energies, but the calibration data shows a 10\%${\sim}$20\% nonlinearity, indicating that the ${\gamma}$ also quenches a little in the crystal, which is also observed in Ref.\cite{CsI(Tl)}. The light yield differences for ${^{241}Am}$ and ${^{137}Cs}$ are taken as systematics, 18.6\% and 15.4\% for $CsI(Na)$ and $CaF_{2}(Eu)$ respectively. Uncertainties from the trigger efficiency are also included in our results.

\subsection{Quality factor of neutron/${\gamma}$ discrimination}

To quantify the discrimination capability between elastic scattering neutrons and ${\gamma}$ events, a quality factor~\cite{Gaitskell} is defined  as
\begin{eqnarray}
K {\equiv} \frac{{\beta}({1-\beta})}{({\alpha}-{\beta})^{2}}
\label{Eq:quality}
\end{eqnarray}
${\alpha}$ means the fraction of signals  passing the selection criteria and ${\beta}$ is the fraction of backgrounds passing the same criteria. For an ideal detector, ${\alpha}$ = 1 and  ${\beta}$ = 0. Therefore, a smaller quality factor means a better discrimination between the signal and background events.
\par
\begin{figure}[H]
\centering
\includegraphics[width=4.3cm,height=6.5cm,angle=270]{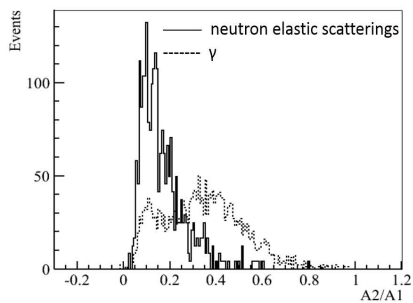}
\caption{$A_{2}/A_{1}$ distribution of $CsI(Na)$ triggered with ND3.}
\label{Fig:separation parameters}
\end{figure}
\begin{figure}[H]
\centering
\includegraphics[width=6.5cm,height=4.3cm]{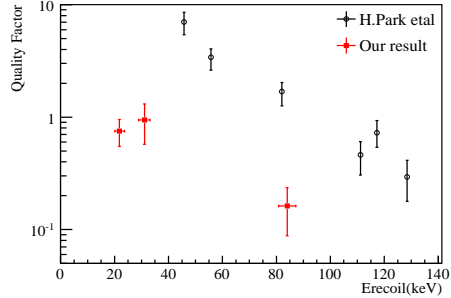}
\caption{Quality factors of $CsI(Na)$. The present results are compared to H.Park etal~\cite{KIMS}. The errors are only statistical. }
\label{Quality factor of CsI(Na)}
\end{figure}
\begin{figure}[H]
\centering
\includegraphics[width=4.3cm, height=6.5cm,angle=270]{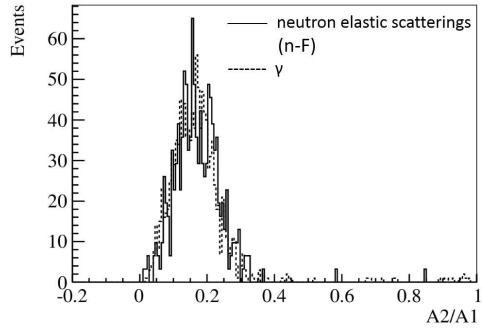}
\caption{$A_{2}$/$A_{1}$ distribution of $CaF_{2}(Eu)$ triggered with ND3.}
\label{Fig:A2/A1 distribution of CaF2}
\end{figure}

A variable $A_{2}$/$A_{1}$~\cite{f90} is defined to calculate the quality factor, where $A_{2}$ is the charge of the first 25~ns of a pulse and $A_{1}$ is the total charge of the pulse. Fig.~\ref{Fig:separation parameters} presents the $A_{2}/A_{1}$ distribution of the ${CsI(Na)}$ crystal and the quality factors at different energies are obtained (Fig.~\ref{Quality factor of CsI(Na)}). The result here is better than the previous test, where the quality factor is calculated with the mean time of a pulse~\cite{KIMS}. For $CaF_{2}(Eu)$, the distribution of $A_{2}$/$A_{1}$ is shown in Fig.~\ref{Fig:A2/A1 distribution of CaF2}, it is hard to discriminate elastic scattering neutron events from ${\gamma}$ events.

\section{Conclusion}
\par
The nuclear recoils of $CsI(Na)$ and $CaF_{2}(Eu)$ crystals are studied with the 14.7~MeV neutron beam. Quenching factors are reported and consistent with the previous work. The quality factor between elastic scattering neutrons and ${\gamma}$ events are obtained for Cs or I in $CsI(Na)$ at various recoil energies, and improved results are obtained by using the new discrimination parameter.
\par
The results indicate that $CsI(Na)$ can discriminate elastic scattering neutrons and ${\gamma}$ backgrounds at a certain extent. While $CaF_{2}(Eu)$ do not have enough capability for neutron/${\gamma}$ discrimination by using A2/A1 at low energy. The systematic uncertainties of quenching factor mainly come from the nonlinearity of the ${\gamma}$ energy response of the crystals. Calibrations of the nonlinearity must be done to improve the measurement accuracy. To extend the measurements to lower nuclear recoil energies, crystals with higher light yield should be used.

\section*{Acknowledgements}
\par
This work is supported by the Ministry of Science and Technology of the People's Republic of China (No.~2010CB833003). We thank L.~Hou, H.T.~Chen and F.~Zhao of CIAE for their help during the experiment and many thanks to T.~Alexander of Pacific Northwest National Lab for his help in editing the language.



\end{document}